\documentclass[sigconf]{acmart}

\newif\ifanon\anontrue

\AtBeginDocument{%
  \acmBooktitle{KEIR Workshop @ CIKM 2026}%
}
\renewcommand\footnotetextcopyrightpermission[1]{}
\acmConference[KEIR @ CIKM 2026]{Workshop on Knowledge-Enhanced Information Retrieval}{November 2026}{Rome, Italy}

\usepackage{multirow} 
\usepackage{tikz}
\usetikzlibrary{arrows.meta,positioning}
\pgfdeclarelayer{bg}
\pgfsetlayers{bg,main}
  \usepackage{booktabs}   
  \usepackage{tabularx}   
  \usepackage{ragged2e}   
  \usepackage{pifont}     
  \usepackage{xcolor}     

\newcommand{\cmark}{{\color{green!50!black}\ensuremath{\checkmark}}}
\newcommand{\xmark}{{\color{red!70!black}\ensuremath{\times}}}
\newcommand{\gold}[1]{{\footnotesize\color{black!60}\emph{(gold: #1)}}}
\newcommand{\ans}[1]{\hspace*{1em}{\color{black!60}\textsc{#1}}:~}

\newcommand{\sys}{CacheGraft}

\begin{document}

\title{CacheGraft: Enhancing Retrieval-Augmented Generation \\ via Latent-Space Document Enrichment}

\author{Vihan Lakshman}
\affiliation{\institution{MIT CSAIL}\country{Cambridge, USA}}\email{vihan@mit.edu}

\author{Peter Baile Chen}
\affiliation{\institution{MIT CSAIL}\country{Cambridge, USA}}\email{peterbc@mit.edu}

\author{Zhuohan Gu}
\affiliation{\institution{MIT CSAIL}\country{Cambridge, USA}}\email{zgu15@mit.edu}

\author{Michael Cafarella}
\affiliation{\institution{MIT CSAIL}\country{Cambridge, USA}}\email{michjc@csail.mit.edu}

\begin{abstract}
Recent work in machine learning has demonstrated that transformer-based language model agents can communicate state not only through text, but also directly in latent space via key-value (KV) caches. In this paper, we argue that such cache-to-cache translation can provide utility beyond the original motivating application of real-time agent communication and can instead be utilized for retrieval-augmented generation (RAG). Building on the Cache-to-Cache (C2C) fusion framework, our proposed system CacheGraft encodes each corpus passage offline into a fused KV cache representation between a larger sharer model (4B parameters in our experiments) and a smaller receiver model (0.6B parameters) such that this cache can be injected into the receiver at query time with no need for the larger model. Across five QA benchmarks and two independent locally-run judges, CacheGraft improves answer quality over the text-only receiver baseline in 29 out of 30 comparisons, of which 17 of the 30 are statistically significant. Notably, these gains are zero-shot with respect to the fuser, which was never trained on retrieval data. These initial results suggest that latent-space document enrichment may be a promising avenue for enhancing the performance of RAG systems, particularly in compute-constrained settings where it is only feasible to deploy a small model at inference time.
\end{abstract}

\ccsdesc[500]{Information systems~Information retrieval}
\keywords{retrieval-augmented generation, KV caching, model collaboration, cache translation}

\maketitle

\section{Introduction}

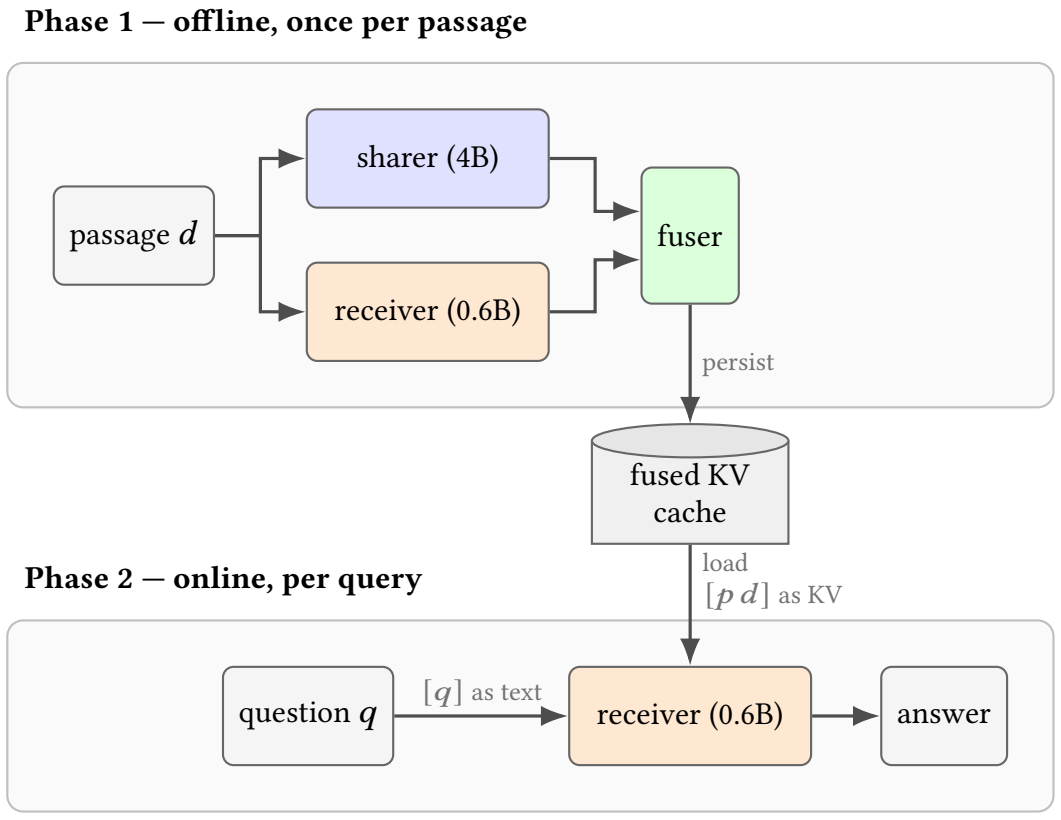
\begin{figure}[t]
\centering
\begin{tikzpicture}[
  font=\scriptsize,
  box/.style={draw=black!60, rounded corners=2pt, minimum height=6.5mm, align=center, inner sep=3pt},
  data/.style={box, fill=black!4},
  big/.style={box, fill=blue!12, minimum width=16mm},
  small/.style={box, fill=orange!18, minimum width=16mm},
  fuse/.style={box, fill=green!14, minimum height=9mm},
  arr/.style={-{Latex[length=1.8mm]}, semithick, black!70},
  phaselab/.style={font=\bfseries\scriptsize, anchor=south west},
  note/.style={font=\tiny, black!55}
]
 
\node[big] (sharer) at (0,0) {sharer (4B)};
\node[small, below=3.5mm of sharer] (recvA) {receiver (0.6B)};
\node[data, left=6mm of sharer.west, anchor=east, yshift=-5.1mm] (psg) {passage $d$};
\node[fuse, right=6mm of sharer.east, anchor=west, yshift=-5.1mm] (fuser) {fuser};
 
\draw[arr] (psg.east) -- ++(3mm,0) |- (sharer.west);
\draw[arr] (psg.east) -- ++(3mm,0) |- (recvA.west);
\draw[arr] (sharer.east) -- ++(3mm,0) |- ([yshift=1.6mm]fuser.west);
\draw[arr] (recvA.east) -- ++(3mm,0) |- ([yshift=-1.6mm]fuser.west);
 
\node[box, fill=black!6, below=9mm of fuser, minimum width=13mm,
      rounded corners=0pt, inner ysep=4pt] (disk) {fused KV\\cache};
\fill[black!10] (disk.north) ellipse [x radius=6.5mm, y radius=1.1mm];
\draw[black!60]  (disk.north) ellipse [x radius=6.5mm, y radius=1.1mm];
\draw[arr] (fuser.south) -- node[note, right, inner sep=2pt] {persist}
      ([yshift=1.1mm]disk.north);
 
\node[small, below=8mm of disk] (recvB) {receiver (0.6B)};
\node[data, left=11.5mm of recvB] (q) {question $q$};
\node[data, right=4.5mm of recvB] (ans) {answer};
 
\draw[arr] (q) -- node[note, above, inner sep=1.5pt] {$[q]$ as text} (recvB);
\draw[arr] (recvB) -- (ans);
\draw[arr] (disk.south) -- node[note, right, inner sep=2pt, pos=0.3, align=left]
      {load\\ $[p\,d]$ as KV} (recvB.north);
 
\begin{pgfonlayer}{bg}
  \draw[black!25, fill=black!2, rounded corners=3pt]
    ([xshift=-3mm, yshift=3mm] psg.west |- sharer.north)
      rectangle ([xshift=3mm, yshift=-3mm] ans.east |- recvA.south);
  \draw[black!25, fill=black!2, rounded corners=3pt]
    ([xshift=-3mm, yshift=3mm] psg.west |- recvB.north)
      rectangle ([xshift=3mm, yshift=-3mm] ans.east |- recvB.south);
\end{pgfonlayer}
 
\node[phaselab] at ([xshift=-3mm, yshift=3.5mm] psg.west |- sharer.north)
      {Phase 1 --- offline, once per passage};
\node[phaselab] at ([xshift=-3mm, yshift=3.5mm] psg.west |- recvB.north)
      {Phase 2 --- online, per query};
\node[note, anchor=north west] at ([xshift=-3mm, yshift=-3.8mm] psg.west |- recvA.south)
      {};
\node[note, anchor=north west] at ([xshift=-3mm, yshift=-3.8mm] psg.west |- recvB.south)
        {};
 
\end{tikzpicture}
\caption{The sharer and fuser run only offline, once per corpus passage, with no
knowledge of any query. The fused KV cache is the only artifact that crosses the
offline/online boundary: at query time the receiver loads it and decodes alone,
reading the passage $[p\,d]$ in latent space and the question $[q]$ as text.}
\label{fig:pipeline}
\end{figure}

Recent breakthroughs in the machine learning community~\cite{fu2026cache, dery2026latent, rossi2026latent} have demonstrated that transformer-based large language model (LLM) agents can communicate information between one another not only through the standard medium of text, but also in \emph{latent space} by translating the key-value (KV) cache from a \emph{sender} model to a \emph{receiver} model's internal vector space. Notably, this translation, typically accomplished by learning a separate adapter model, is feasible even if the two models have a differing number of weights and layers~\cite{fu2026cache, dery2026latent, rossi2026latent}.  

The overarching motivation driving this effort in KV cache translation is the vision of heterogeneous LLMs collaborating in a real-time multi-agent system by passing state via KV cache vectors as opposed to raw text. By operating in latent space, KV cache translation offers the promise of improving both the \emph{quality} of multi-agent computation, since KV caches constitute richer conceptual representations of information in an LLM's native language, and the \emph{performance} of such systems by eliminating the model's initial step of converting a textual input into KV vectors. 

The central observation behind this paper is that KV cache translation can be useful beyond the original motivating goal of real-time multi-agent LLM communication. To that end, we introduce the \sys{} framework applying the KV cache translation paradigm for retrieval-augmented generation (RAG) where an LLM marshals knowledge from an external corpus to enhance its context before generating a response. Applying KV cache translation to RAG is not immediately obvious because typical RAG deployments often consist of a single LLM; it is unclear where cross-model cache translation even comes into the picture. Our key insight is that we can leverage KV cache translation during the \emph{offline indexing} phase of RAG pipelines by enriching the representation of a document through fusing the KV cache of a larger model (that cannot be deployed at query time) with the cache of a smaller model that does run online. Thus, the larger model serves as a de facto teacher and guides the smaller receiver model to better ``read" the contents of the documents. 

Unlike multi-agent communication, the offline phase of a RAG pipeline is not latency-critical, making the overhead of KV cache fusion more palatable than in the real-time communication setting. Moreover, this offline phase may enable using more powerful, but computationally expensive LLM models for indexing. These insights drive the development of our \sys{} framework. In summary, we make the following contributions in this paper:
\begin{itemize}
\item \textbf{Conceptual Insight:} To our knowledge, this work is the first to make the observation that KV cache translation can be leveraged in applications outside of the canonical motivating example of real-time agent-to-agent communication. In particular, we demonstrate that the Cache-to-Cache (C2C) fusion technique~\cite{fu2026cache} can lead to improved generation quality in retrieval-augmented generation (RAG) settings. 


\item \textbf{Proof-of-Concept Evaluation:} We instantiate \sys{} using an off-the-shelf pretrained KV cache fusion model published by the C2C team\footnote{\url{https://huggingface.co/nics-efc/C2C_Fuser}} that fuses the KV cache of Qwen3-4B-Base model with the cache from a Qwen3-0.6B model into the latter's KV space. We evaluate our proposed system\footnote{Our code is available at \url{https://github.com/vihan-lakshman/CacheGraft}} on five knowledge-intensive RAG question-answering benchmarks which differ from the original C2C evaluation benchmarks which largely focus on reasoning tasks. Interestingly, \sys{} achieves these gains using a cache fusion model that was not explicitly trained for these retrieval applications. 

\item \textbf{Long-Term Research Vision:} In addition to our proof-of-concept experiment results, we argue that KV cache fusion for RAG applications is a rich and understudied area and we highlight several open research problems for the community. 

\end{itemize}

\section{Related Work}

Our work in this paper sits at the confluence of three research threads in machine learning and information retrieval that were previously disconnected. First is the aforementioned line of work on KV cache translation~\cite{fu2026cache, dery2026latent, rossi2026latent} including the C2C framework~\cite{fu2026cache} we directly use in our experiments. We focus on using C2C in our evaluation of CacheGraft because, to our knowledge, it has demonstrated efficacy with the largest gap between the sharer and receiver models (4B to 0.6B parameters) and thus constitutes the most scalable translation method to date. 

Secondly, on the RAG front, a number of works have proposed precomputing and storing a model's KV cache representation of documents and using the cache, not the raw text, at query time to reduce latency by eliminating the step of converting text into KV values~\cite{lu2025turborag, chen2026log, yao2025cacheblend, ma2025block}. However, these methods all use the KV cache of a single model and do not consider enriching the cache representation with signal from an additional, more powerful LLM, which differentiates our approach in this paper. 

Finally, multiple research efforts have also examined the problem of \emph{document enrichment}, namely preprocessing the text in a corpus to be more amenable to RAG applications~\cite{chen2025enrichindex, o2026autoindex, uzan2026document, lei2026rl}. \sys{} can also be viewed as a document enrichment strategy, but in latent space as opposed to at the textual level. 


\section{Background: Cache-to-Cache (C2C)}

In this section, we briefly review the details of the C2C framework~\cite{fu2026cache} as \sys{} repurposes this same fusion model for offline document enrichment. In short, C2C enables a \emph{sharer} LLM to pass
contextual understanding to a \emph{receiver} LLM directly in latent space, by fusing the sharer's key-value (KV) cache into the receiver's. Both
underlying models remain frozen; the only trained component is a lightweight
per-layer \emph{fuser} network. More formally, let $x = [x_0, \dots, x_{n-1}]$ denote the input token sequence (in our
setting, the RAG prompt; see Section~\ref{sec:cachegraft}). The sequence is
tokenized independently by each model. Both frozen models then prefill on the same input: the receiver
produces its per-layer KV cache $\mathcal{C}_n(x)$ for layers
$n = 1, \dots, N$, and the sharer produces its cache
$\mathcal{C}^{\mathcal{S}}_m(x)$ for its own layers $m = 1, \dots, N^\prime$. The text input is consumed twice, once by each
model's prefill.

Because the two models may differ in depth, a layer-mapping strategy
$\mathcal{G}$ assigns to each receiver layer $n$ a sharer layer
$\mathcal{G}(n)$; C2C adopts \emph{terminal alignment}, pairing final layers
first and proceeding backward. A per-layer fuser $\mathcal{F}_n$ then produces the fused cache:
\begin{equation}
    \mathcal{C}^{\mathcal{F}}(x)
    \;=\;
    \bigl\{\,
        \mathcal{F}_n\bigl(\mathcal{C}_n(x),\,
        \mathcal{C}^{\mathcal{S}}_{\mathcal{G}(n)}(x)\bigr)
    \,\bigr\}_{n=1}^{N}.
    \label{eq:fusion}
\end{equation}
The fused cache $\mathcal{C}^{\mathcal{F}}(x)$ lives entirely in the
receiver's KV space, has the same sequence length $n$ as the input (no extra
token positions are introduced), and \emph{replaces} the receiver's original
prompt cache.

At inference time, the receiver performs text generation alone. With generated prefix
$y_{[0:i]} = [y_0, \dots, y_{i-1}]$ and the receiver's ordinary
self-produced cache $\mathcal{C}(y_{[0:i]})$ over that prefix, the next token
is predicted as
\begin{equation}
    y_{i+1}
    \;=\;
    \mathcal{P}\bigl(y_i \,\big|\,
        \mathcal{C}^{\mathcal{F}}(x) \oplus \mathcal{C}(y_{[0:i]})\bigr),
    \label{eq:decode}
\end{equation}
where $\oplus$ denotes sequence-wise concatenation. After fusion, the sharer
plays no further role: newly generated tokens append receiver-native cache
entries behind the fused prompt cache exactly as in standard autoregressive
decoding.
 
Two properties of this construction are essential for the RAG setting in \sys{}. First, the fused cache is a self-contained artifact: Equation~\ref{eq:decode}
requires only $\mathcal{C}^{\mathcal{F}}(x)$ and the receiver, not the sharer
or the fuser. Second, with causal attention, the fused entries at a given
position depend only on tokens at earlier positions. Together these
properties allow the fusion step of Equation~\ref{eq:fusion} to be moved
offline entirely, computed once per document without knowledge of any query.

One limitation of the C2C framework in the interactive multi-agent setting is that it requires running a forward pass of the text input $x$ through \emph{both} the sharer and receiver model, which limits the total possible amount of latency savings. A major goal of subsequent work in KV cache translation, such as Latent Cache Flow~\cite{rossi2026latent}, is to eliminate this shortcoming. However, in the RAG setting, this fusion approach is not a notable hindrance since the cache enrichment step happens offline and not at query time. 

\begin{table*}[t]
  \centering\scriptsize
  \setlength{\tabcolsep}{2.4pt}
  \caption{Performance of \sys{} in comparison to using only the 0.6B receiver model and only the 4B sharer model across two LLM judges and the contains metric, which measures if the correct answer is included in the output text. \sys{} improvements over the receiver-only baseline are marked in bold. $\Delta$ reports \sys{} minus Receiver-Only for each metric.}
  \label{tab:cross}
  \begin{tabular}{llcc cccc cccc cccc}
  \toprule
  & & & & \multicolumn{4}{c}{\textbf{phi-4}} & \multicolumn{4}{c}{\textbf{Gemma}} & \multicolumn{4}{c}{\textbf{contains}} \\
  \cmidrule(lr){5-8}\cmidrule(lr){9-12}\cmidrule(lr){13-16}
  Dataset & Setting & $n$ & Top-1 Recall
  & Receiver-Only & \sys{} & Sharer-Only & $\Delta$
  & Receiver-Only & \sys{} & Sharer-Only & $\Delta$
  & Receiver-Only & \sys{} & Sharer-Only & $\Delta$ \\
  \midrule
  \multirow{2}{*}{AdversarialQA} & retrieved & \multirow{2}{*}{416} & 59.4 & 30.8 & \textbf{36.8} & 41.8 & $+6.0^{*}$ & 34.9 & \textbf{38.9} & 48.3 & $+4.1$ & 24.5 & \textbf{30.3} & 33.7 & $+5.8^{*}$ \\
   & oracle &  & --- & 42.1 & \textbf{49.3} & 59.1 & $+7.2^{*}$ & 44.2 & \textbf{51.4} & 63.9 & $+7.2^{*}$ & 33.9 & \textbf{42.5} & 48.3 & $+8.7^{*}$ \\
  \midrule
  \multirow{2}{*}{SQuAD} & retrieved & \multirow{2}{*}{800} & 68.1 & 51.8 & \textbf{59.6} & 66.4 & $+7.9^{*}$ & 59.2 & \textbf{62.5} & 70.8 & $+3.2^{*}$ & 48.6 & \textbf{55.5} & 61.1 & $+6.9^{*}$ \\
   & oracle &  & --- & 73.0 & \textbf{79.4} & 86.5 & $+6.4^{*}$ & 78.6 & \textbf{81.8} & 89.4 & $+3.1$ & 69.9 & \textbf{76.9} & 81.2 & $+7.0^{*}$ \\
  \midrule
  \multirow{2}{*}{CovidQA} & retrieved & \multirow{2}{*}{800} & 53.2 & 33.1 & \textbf{36.9} & 41.9 & $+3.8^{*}$ & 50.8 & \textbf{53.8} & 59.8 & $+3.0$ & 21.0 & \textbf{22.6} & 24.4 & $+1.6$ \\
   & oracle &  & --- & 53.7 & \textbf{59.1} & 61.1 & $+5.4^{*}$ & 71.7 & \textbf{76.9} & 80.0 & $+5.2^{*}$ & 32.6 & \textbf{35.4} & 37.2 & $+2.8$ \\
  \midrule
  \multirow{2}{*}{TriviaQA} & retrieved & \multirow{2}{*}{800} & 54.3 & 38.1 & \textbf{41.6} & 57.8 & $+3.5^{*}$ & 49.2 & \textbf{52.4} & 74.1 & $+3.1^{*}$ & 45.9 & \textbf{48.9} & 65.5 & $+3.0^{*}$ \\
   & oracle &  & --- & 50.8 & \textbf{54.0} & 70.1 & $+3.2$ & 62.9 & \textbf{63.4} & 85.0 & $+0.5$ & 61.0 & \textbf{62.8} & 81.5 & $+1.8$ \\
  \midrule
  \multirow{2}{*}{QASPER} & retrieved & \multirow{2}{*}{583} & 36.2 & 15.3 & \textbf{15.7} & 17.2 & $+0.4$ & 35.9 & 33.9 & 35.5 & $-2.0$ & 17.7 & \textbf{18.8} & 18.2 & $+1.1$ \\
   & oracle &  & --- & 34.7 & \textbf{37.5} & 37.9 & $+2.9$ & 57.4 & \textbf{60.8} & 59.2 & $+3.4$ & 32.7 & \textbf{39.0} & 34.8 & $+6.3^{*}$ \\
  \bottomrule
  \end{tabular}\\[2pt]
  \footnotesize{$^{*}$ $p<0.05$, exact two-sided paired McNemar over questions.}
  \end{table*}
  
\section{CacheGraft: Adapting C2C for RAG}
\label{sec:cachegraft}

We now describe the details of our \sys{} system. As a matter of notation, we will denote a RAG prompt by $x = [p, d, q]$ which is a concatenation of a fixed preamble $p$ (also known as a system prompt), a retrieved document $d$, and the question $q$. We focus our attention on knowledge-intensive question-answering applications that require a factual answer typically not found in the model's own weights. 

With causal attention, both models' pre-fusion cache entries at
positions $i \le |p\,d|$ are independent of any token at position $j > i$; the projector acts
position-wise and layer-wise on those entries; and the fusion of $[p\,d]$ does not depend
on $q$. Hence computing $[p,d,q]$ in one pass and computing $[p,d]$ alone yield identical KV over
$[p,d]$. As a consequence, for top-1 retrieval with a fixed preamble, the document always occupies the same
position range, so no RoPE re-rotation~\cite{chen2026log} is needed. We leave extending our framework to support top-$k$ retrieval for $k > 1$ using techniques such as TurboRag finetuning~\cite{lu2025turborag} or CacheBlend~\cite{yao2025cacheblend} for future work.  

Figure~\ref{fig:pipeline} depicts the full \sys{} pipeline. Offline, for each passage, we prefill $[p,d]$ with fusion enabled and
persist the receiver-space KV, keyed by a content hash. We also maintain an additional vector database index of document embeddings. Online, we identify the most relevant document to the query based on embedding similarity and load that document's fused KV cache into the
\emph{base} receiver and decode the question. Crucially, we do not need anything from the sharer or the cache projection fuser model on the query path.

\section{Experiments}
\label{sec:experiments}

\subsection{Experimental Setup}

We conduct all experiments on a single NVIDIA A100-40GB GPU and use standard greedy decoding at inference time. 

\paragraph{Models.} In all reported experiments, we use Qwen3-0.6B~\cite{yang2025qwen3} as our receiver and Qwen3-4B-Base~\cite{yang2025qwen3} as the sharer, with the released C2C projector for that pair. The projector is trained on general chat distillation and never sees
retrieval data, so every result is zero-shot with respect to fuser training.

\paragraph{Data.} We utilize five standard knowledge-intensive question answering benchmarks for our RAG evaluation: SQuAD v1.1~\cite{rajpurkar2016squad}, AdversarialQA~\cite{bartolo2020beat},
CovidQA~\cite{moller2020covid}, QASPER~\cite{dasigi2021dataset} and
TriviaQA~\cite{joshi2017triviaqa}. For search, we use bge-small~\cite{bge_embedding} embeddings to perform dense top-1 retrieval. For evaluation, we retain one question per gold passage. We also run our experiments under \texttt{retrieved} and \texttt{oracle} (perfect-retrieval) conditions to understand the impact of search quality on the overall RAG performance.

\paragraph{Metrics.} We utilize the LLM-as-judge mechanism~\cite{zheng2023judging} as our primary evaluation metric. We evaluate all RAG outputs using two judges, both locally-run open-weight models scored on raw
unmodified generations: \texttt{phi-4}~\cite{abdin2024phi} and
\texttt{Gemma-2-9B}~\cite{team2024gemma}. We also report \texttt{contains} which measures whether the generated answer contains the gold ground truth answer included in the test dataset. We also compute statistical significance measures using the exact two-sided paired McNemar test over the set of evaluation questions. We do not rely on traditional RAG metrics such as Exact Match (EM) and F1 score because we observed that the generated answers from fused KV caches tend to be more verbose (illustrated in Table~\ref{tab:examples}) which hurts performance on these token-centric metrics. 

\subsection{Results}
\label{sec:results}


\begin{table}[t]                                    
    \centering\scriptsize                                                    
    \caption{Example outputs of \sys{} and the receiver-only Text-RAG baseline.  
    The last two rows are failure cases: in each, \sys{} answers with a different span of the same passage. }                               
    \label{tab:examples}                                             
    \begin{tabularx}{\columnwidth}{@{}>{\raggedright\arraybackslash}X@{}}     
      \toprule                                   
      \addlinespace[1pt]                         
      Over how many species of trees can be found in the Saxon Garden? \gold{100} \\                                                                                                                                                          
      \addlinespace[3pt]                     
      \ans{text-RAG} 15.5\enspace\xmark \\[3pt]                                    
      \ans{\sys{}} 100\enspace\cmark \\           
      \addlinespace[4pt]                             
      \midrule                                      
      \addlinespace[3pt]                              
      Where was the Continental Edison Company located? \gold{France} \\           
      \addlinespace[3pt]                                                                                                                                                                                    
      \ans{text-RAG} New York City.\enspace\xmark \\[3pt]                                                                                                                                                   
      \ans{\sys{}} The Continental Edison Company was located in France.\enspace\cmark \\                                                                                                                   
      \addlinespace[4pt]                                                                                                                                                                                    
      \midrule                                                                                                                                                                                              
      \addlinespace[3pt]                                                                                                                                                                                    
      Who did Rollo sign the treaty of Saint-Clair-sur-Epte with? \gold{King Charles III} \\                                                                                                                
      \addlinespace[3pt]                                                                                                                                                                                    
      \ans{text-RAG} Rollo.\enspace\xmark \\[3pt]                                      
      \ans{\sys{}} Rollo signed the treaty of Saint-Clair-sur-Epte with King Charles III of West Francia.\enspace\cmark \\                                                                                  
      \addlinespace[4pt]                           
      \midrule                                       
      \addlinespace[3pt]                             
      Where were missing patents later found? \gold{patent archives} \\                    
      \addlinespace[3pt]                           
      \ans{text-RAG} The missing patents were later found in patent archives.\enspace\cmark \\[3pt]                             
      \ans{\sys{}} The missing patents were found in the United States, Britain, and Canada.\enspace\xmark \\                                                         
      \addlinespace[4pt]                              
      \midrule                                       
      \addlinespace[3pt]                               
      Whose writings were widely quoted by the Third Reich? \gold{Luther} \\               
      \addlinespace[3pt]                             
      \ans{text-RAG} Luther's writings were widely quoted by the Third Reich.\enspace\cmark \\[3pt]                            
      \ans{\sys{}} The writings of Heinrich Himmler were widely quoted by the Third Reich.\enspace\xmark \\                  
      \addlinespace[1pt]  
      \bottomrule       
    \end{tabularx}                         
  \end{table}

As shown in Table~\ref{tab:cross}, we find that delivering the passage as a projected KV cache improves over using the receiver with raw text on nearly every metric across every benchmark by up to 8 percentage points, with the exception of QASPER, which shows similarly clear gains with an oracle retriever but a disagreement between the two judges when using top-1 embedding-based search.

In Table~\ref{tab:screen}, we observe that QASPER has the smallest gap between the sharer and receiver RAG quality which explains the more mixed results from cache fusion: there is limited headroom for the sharer to lift the receiver. Overall, Table 3 suggests two preconditions for CacheGraft to provide value on knowledge-intensive RAG tasks: the sharer must have headroom over the receiver, and the documents must supply factual signal the sharer can exploit. 


We also include select examples of \sys{}'s successes and failures on the SQuAD dataset in Table~\ref{tab:examples}. We observe that cache fusion can indeed steer the receiver model towards relevant portions of the passage in several cases, but also makes mistakes. We also note that the outputs from \sys{} tend to be more verbose than both the receiver and sharer models which also limits \sys{}'s total latency gains as shown in Table~\ref{tab:cost}. Nevertheless, Table ~\ref{tab:cost} still illustrates an advantage for \sys{} in terms of GPU memory consumption. We posit that the increased verbosity of \sys{}'s fused KV cache can be mitigated by explicitly training for retrieval tasks, which we highlight as another direction for future work. 

\begin{table}[t]
  \centering\small
  \caption{Relationship between CacheGraft's gains and the opportunity for improvement, all computed in the oracle-retrieval setting. Headroom is Sharer-Only minus Receiver-Only under contains. Answer-in-doc is the sharer's \texttt{contains} improvement when given the gold passage versus answering closed-book (no passage). Gain is CacheGraft's improvement over Receiver-Only under the phi-4 judge.}
  \label{tab:screen}
  \begin{tabular}{lccc}
  \toprule
  Dataset & Headroom & Answer-in-doc & Gain \\
  \midrule
  AdversarialQA & $+14.4$ & $+34.6$ & $+7.2$ \\
  SQuAD         & $+11.3$ & $+69.0$ & $+6.4$ \\
  CovidQA       & $+4.4$  & $+34.1$ & $+5.4$ \\
  TriviaQA      & $+20.5$ & $+40.7$ & $+3.2$ \\
  QASPER        & $+2.0$  & $+30.2$ & $+2.9$ \\
  \bottomrule
  \end{tabular}
  \end{table}

\begin{table}[t]
  \centering\small
  \caption{Query-time cost, batch size 1, greedy decoding with 24 maximum new tokens, warm,
  over 100 SQuAD questions on one A100-40GB.}
  \label{tab:cost}
  \begin{tabular}{lccc}
  \toprule
  System & Latency (ms) & GPU Memory (GB) & \# Tokens \\
  \midrule
  Receiver text-RAG (0.6B) & $334$ & $1.22$ & $8.2$ \\
  CacheGraft (0.6B + Cache) & $574$ & 1.27 & $14.3$ \\
  Sharer text-RAG (4B) & $652$ & $7.62$ & $12.6$ \\
  \bottomrule
  \end{tabular}
  \end{table}



\section{Limitations and Open Problems}

\subsection{Limitations}
The primary practical limitation of our proposal is storage since we must store the KV cache for every document which can consume orders of magnitude more space than raw text (on the order of a terabyte for 100K documents, absent any optimizations). This problem is inherited from prior works that propose to precompute the KV caches for retrieval such as TurboRAG~\cite{lu2025turborag}, Block Attention~\cite{ma2025block}, and CAG~\cite{chan2025don} and can be mitigated with KV cache compression techniques~\cite{shi2024keep} that are complementary to our work. Another limitation of our present study is that we focus on retrieving a single document and only consider a single C2C fuser model in our evaluations. Extending our system to support top-$k$ retrieval and additional models is another promising direction for future work. 

\subsection{Open Problems}
\label{sec:open-probs}
The results in this paper present a proof of concept demonstration of KV cache fusion to enhance RAG applications and open many directions for future research in this domain. Below, we highlight two particularly interesting directions:

\begin{enumerate}

\item \textbf{Training custom fuser models tailored for RAG:} As mentioned previously, we consider a single off-the-shelf C2C fuser model that was not targeted for retrieval tasks. Could we improve the performance of \sys{} by training a fuser model with retrieval-specific training objectives and datasets? How far can we scale the gap between the sharer and receiver model before cache fusion fails? Could we even train the fuser to be more compressible to augment standard KV cache compression techniques?

\item \textbf{Combining with textual document enrichment:} Prior work such as EnrichIndex~\cite{chen2025enrichindex} and AutoIndex~\cite{o2026autoindex} propose to modify document text to improve RAG performance. How does text-level document enrichment interplay with the latent space strategy of our method? Is there a text enrichment method that enhances KV cache translation beyond simply using the raw documents?
\end{enumerate}
\section{Conclusion}

Thus far, the idea of translating the KV cache from one LLM to another possibly very different model has been applied exclusively to communicating state in real-time multi-agent systems. The key insight at the heart of this paper is that the same idea can be repurposed for RAG by applying KV cache translation on documents offline between a larger sharer model and a smaller receiver to enhance the receiver's online answer quality. In fact, we argue that RAG might be an even \emph{better} fit for the Cache-to-Cache fusion paradigm because running forward passes through both the sharer and receiver model (as C2C requires) is a negligible cost for offline document indexing, but may be limiting in real-time inference settings. We operationalize this insight in the form of \sys{}, a prototype RAG system that utilizes C2C fusion to enrich a small receiver model's KV cache representation with signal from a larger model that does not run at query time. Our initial evaluation extends C2C to zero-shot knowledge-intensive question answering tasks not present in fuser training where we find that \sys{} consistently improves generation quality over the receiver-only text baseline on datasets where the larger sharer model has headroom to transfer. Perhaps surprisingly, these positive results on question-answering come from using an off-the-shelf pre-trained C2C fuser model that was not tuned for retrieval in any respect. Our findings also raise a number of interesting questions for future work, which we outline in Section~\ref{sec:open-probs}, that we hope will drive further work on latent-space document enrichment for RAG applications. 


\bibliographystyle{plain}
\bibliography{custom}

\end{document}